# A Search for Point Sources of Cosmic Primary Rays Which Produce Single Muon Tracks at Ground Level

A. Roesch, J. Carpenter, S. Desch, J. Gress, T. F. Lin, and J. Poirier
*Physics Department, 225 NSH, University of Notre Dame, Notre Dame, IN 46556, USA*

## Abstract

An increased data sample of identified secondary muons is collected at detection level. Project GRAND identifies secondary cosmic ray muons from electrons to 96% precision utilizing a thin steel absorber and tracking (PWC) chambers (Gress et al., 1991). Since single (non-shower) tracks are 75% muons, the number of misidentified muons in this sample is ~1%. The angles of these identified muons are measured to an average projected angle precision of 0.26 degrees. The resulting angular resolution for the primary is about ± 5 degrees (in east and north projections). Since there is a 1.5% probability that a 100 GeV gamma primary will produce a single muon track at ground level, the high statistics allow a search for angular enhancements due to uncharged particles (gammas) which are not deflected in the galactic magnetic field. More than 100 billion muons are identified and analyzed in right ascension and declination directions from data taken during the last two years. These results are compared with similar earlier data. A table of stellar sources is examined.

## 1 Introduction:

Many objects outside our solar system emit cosmic rays which, upon hitting the earth's atmosphere, interact with atmospheric nuclei and produce pions. These pions can decay into muons and neutrinos in competition with their hadronic interactions. Most of the muons at the detection level of Project GRAND at Notre Dame are produced by hadronic particles. There is no known method of determining the original source of these charged particles because their paths have been deflected by magnetic fields. Fortunately, about 1.5 percent of neutral gamma rays at 100 GeV also produce muons that can be detected by Project GRAND (Fasso & Poirier, 1999). If the muons which originated from gamma rays can be seen above background, the objects which produced the gamma rays are seen as point sources at a particular right ascension and declination.

Forty point sources, taken from a 1990 list of pulsars, binary x-rays, and other possible point sources generated by the Cygnus group (Lu et al., 1990) were studied. Data on possible detection of the point sources by Project GRAND of Notre Dame were first published in the Proceedings of the 1991 International Cosmic Ray Conference (Poirier et al., 1991). Additional data have been taken and presented at subsequent International Cosmic Ray Conferences (Trzupek et al., 1993; Poirier et al., 1997). Since the original paper, Project GRAND has undergone continuing improvements. It currently operates with 64 stations (compared to the original 16). Besides the advantage of adding more stations, each station has been improved to be more reliable, and therefore more days of data can be taken. The data acquisition systems have also been improved. New software and data storage systems ensure that more data are taken for each day; in general, better data can be taken on a more regular basis. Thus the quality of the results of this point source study increases each two years due to the constantly improving efficiency of the detectors and the data taking process of Project GRAND.

## 2 Experiment:

Secondary cosmic rays are composed mainly of muons, electrons, and gamma rays. In order to study the number of muons produced by primary gamma rays emitted from point sources, it is necessary to differentiate between muons and other particles. This can be done with the detectors used by Project GRAND (Gamma Ray Astrophysics at Notre Dame) located at 41.7 degrees North latitude and 86.2 degrees

West longitude, 220 meters above sea level, located just north of the University of Notre Dame campus in Indiana. GRAND is an array of 64 stations of detectors placed in wood huts spaced 14 meters apart in an array of 100 m x 100 m. A 50 mm steel plate between the third and fourth of the four detectors in each hut is used to determine which particles being detected are muons (Gress et al., 1991). Each set of detectors is capable of differentiating muons from other particles with 96 percent accuracy, and the angle of the muon track can be measured with an average resolution of 0.26 degrees in each of the two orthogonal planes. However, the angle of the detected secondary muon deviates from the angle of the primary gamma ray that entered the atmosphere. The primary gamma rays have hadronic interactions in the atmosphere which produce pions. These pions then decay into muons, and the muons scatter in the air and deflect in the earth's magnetic field. Each of these processes increases the possible difference between the muon angle and the angle of the original gamma ray. The position of the original point source of the gamma ray can be determined with a ± 5° resolution governed by these various physical processes which degrade this resolution before the tracks reach Project GRAND's detectors.

Data are taken by counting the total number of muons from each point in the sky which is accessible to GRAND, which is 0 to 360 degrees right ascension and -20 to 90 degrees declination. About 1.5 percent of the gamma rays at 100 GeV produce secondary muons which reach detection level. Data are accumulated in units of entire sidereal days. These data are added to the days of 'smooth' data, where the detection efficiency was uniform throughout that day. In this experiment, data from 'smooth' days are summed over a period of two years. This produces high statistics which may be capable of detecting the small number of muons created in the atmosphere from gamma rays originating at these point sources.

## 3 Data and Analysis:

After data were taken over a period of one sidereal day, it was determined whether the data for that day were 'smooth'. Because of the rotation of the earth, different right ascension angles were measured at different times of the day. If a detector was not recording muons for some time during the day, the data would show a drop in the muons with right ascension angles near GRAND's zenith angle at that time. For example, if 60 huts were operational throughout the day, the counts from that data would accurately depict the muon count from each right ascension angle relative to the other right ascension angles. However, if some of those huts were not operational for a few hours of that day, there would be fewer muons at right ascension angles which were at GRAND's zenith angle during those hours. A possible signal causing an increase in muons at a particular right ascension and declination is expected to be a small effect compared to the background. Even a small drop in muons for any right ascension angles could be confused as the background next to a peak. Since this non-smooth data would not represent what was emitted from the point source, they were not used in the results that follow. The process of determing which data were 'smooth' involved summing the day's data for each right ascension angle. If the muon count did not vary too much over the spectrum of right ascension angles, the data were considered 'smooth'. This was determined by adding the counts for each right ascension angle regardless of declination. Different right ascension angles represented the maximum number of counts and the minimum number of counts. In order for the day's data to be considered smooth, the difference between the maximum count and the minimum count was required to be less than ± 5 percent from the average of the counts from all right ascension angles. All smooth data were added to obtain the number of muon counts at each right ascension and declination for the past two years.

At this point, the background of muon counts was subtracted from the counts at the indicated point source location to determine the signal of single-track muons from that source. The resolution of measuring the point source primary gamma ray which produced a single track muon is ± 5°. Thus the muon count from a point source was measured by adding the counts in a square of ± 5° declination angle($\theta$) and ± 5° / cos($\theta$) right ascension angle. Division by the cosine of the declination angle is used because, at larger declination angles (smaller values of cos($\theta$)), a greater span of right ascension angles have to be used to measure the ± 5° resolution. Background was determined by averaging the counts of two squares of data that were of the same size and adjacent to the position as the original square, taken on either side of the point source square in right ascension. Squares could be taken from different right ascension angles, since muon counts vary insignificantly with changes in right ascension. Background was not taken at nearby declination angles because the counts vary rapidly with declination angle (Poirier et al., 1997).

Table 2 contains the results for the forty different point sources chosen. Included in Table 2 is the number of the point source refering to its position on the list of pulsars, binary x-rays, and other point sources originally studied by the Cygnus group (Lu et al., 1990). Table 2 contains the declination and right ascension angles of the point source, number of muon counts from the point source (sig) and its error (dsig), the background muon counts in parts per million (back), the ratio of signal to background in parts per 1000 (flux), and the ratio of the signal to its error (sig/dsig).

## 4 Conclusions:

Figure 1 is a plot of the number of point sources producing signal to error ratios in ranges of 0.5 units. This plot was fitted with two exponential functions. The function with two parameters is shown with the dotted line and the function with three parameters is shown with the solid line; the plot with the added variable "C" allows for a possible positive excess of sources seen by GRAND. The values and their errors were fit by a maximum likelihood function; the errors were determined including the correlations between the constants. Table 1 lists the values of these parameters that produced the best fit to the data.

Ten of the forty point sources have signal to error ratios of one or above and two have two or above. Of the forty sources searched, twenty-six of the ratios were positive, eleven negative, and three zero. The sum of the 40 signal to error ratios is +14.6.

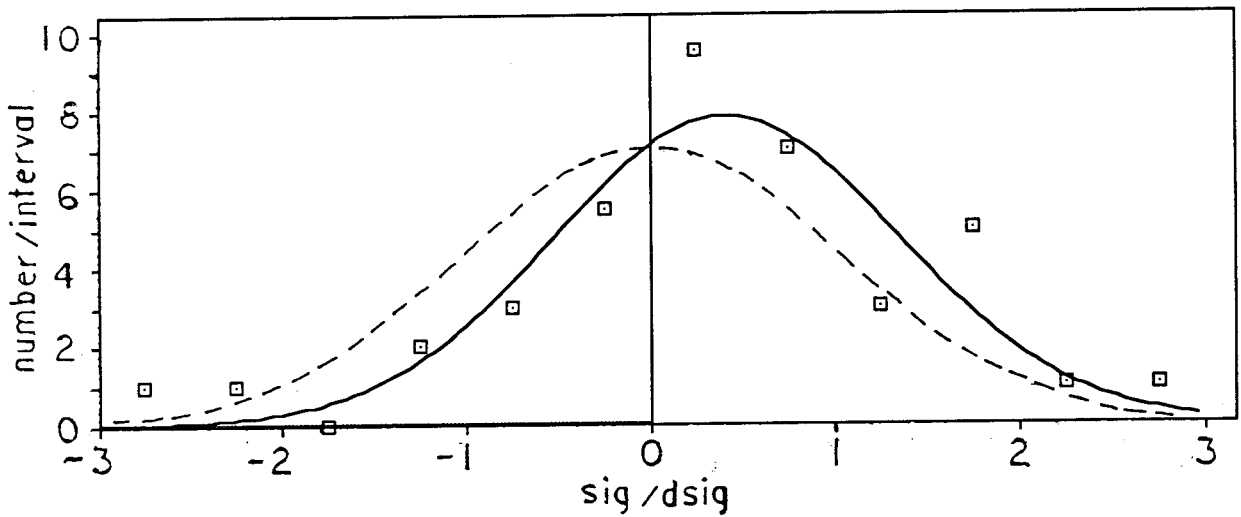

**Figure 1:** A graph of the number of point sources which were found to have signal to error ratios in intervals 0.5 units wide. The horizontal axis is the value of the signal to error ratio, expressed in Table 2 as sig/dsig. The vertical axis is the number of times the ratio occurred in intervals of 0.5. Table 1 contains the best fit parameters of two exponential curve fits to this histogram.

Table 1: Coefficients of the function A*exp{-(x-C)$^2$/(2B$^2$)} which were fit to the data

|   | A*exp{-(x-C)$^2$/(2B$^2$)} | A*exp{-(x)$^2$/(2B$^2$)} |
|---|---|---|
| A | 7.84 ± 0.58 | 7.08 ± 0.46 |
| B | 1.11 ± 0.12 | 1.12 ± 0.13 |
| C | 0.36 ± 0.18 | |

Table 2: Project GRAND's results from examining 40 pulsars, binary x-rays, and other sources

| no. | DEC | RA | sig | dsig | back | flux | sig/dsig |
|---|---|---|---|---|---|---|---|
| 1 | 3 | 285 | 20466 | 9721 | 63.0 | 0.3 | 2.1 |
| 2 | 5 | 287 | 973 | 10548 | 74.2 | 0.0 | 0.1 |
| 3 | 5 | 279 | 9251 | 10546 | 74.1 | 0.1 | 0.9 |
| 4 | 8 | 148 | -33963 | 11859 | 93.8 | -0.4 | -2.9 |
| 5 | 9 | 94 | 8015 | 12266 | 100.3 | 0.1 | 0.7 |
| 6 | 10 | 287 | -4882 | 12748 | 108.4 | 0.0 | -0.4 |
| 7 | 10 | 292 | -2282 | 12749 | 108.4 | 0.0 | -0.2 |
| 8 | 12 | 187 | 2781 | 13460 | 120.8 | 0.0 | 0.2 |
| 9 | 14 | 104 | 12112 | 14353 | 137.3 | 0.1 | 0.8 |
| 10 | 15 | 290 | 2739 | 14707 | 144.2 | 0.0 | 0.2 |
| 11 | 18 | 98 | 12475 | 15812 | 166.7 | 0.1 | 0.8 |
| 12 | 20 | 299 | -10907 | 16648 | 184.8 | -0.1 | -0.7 |
| 13 | 21 | 294 | -433 | 16868 | 189.7 | 0.0 | 0.0 |
| 15 | 22 | 83 | -34794 | 17219 | 197.7 | -0.2 | -2.0 |
| 17 | 26 | 84 | -20174 | 18403 | 225.8 | -0.1 | -1.1 |
| 18 | 28 | 194 | -3433 | 19006 | 240.8 | 0.0 | -0.2 |
| 19 | 29 | 299 | 9120 | 19324 | 249.0 | 0.0 | 0.5 |
| 20 | 29 | 298 | 4639 | 19324 | 249.0 | 0.0 | 0.2 |
| 21 | 31 | 58 | -3499 | 19495 | 253.4 | 0.0 | -0.2 |
| 22 | 32 | 299 | 4420 | 19849 | 262.7 | 0.0 | 0.2 |
| 23 | 32 | 36 | -855 | 19825 | 262.0 | 0.0 | 0.0 |
| 25 | 35 | 299 | 8137 | 21853 | 318.4 | 0.0 | 0.4 |
| 26 | 35 | 254 | 24636 | 21837 | 317.9 | 0.1 | 1.1 |
| 27 | 36 | 305 | 31017 | 21987 | 322.3 | 0.1 | 1.4 |
| 28 | 37 | 273 | 6555 | 22102 | 325.7 | 0.0 | 0.3 |
| 29 | 38 | 326 | 38858 | 21988 | 322.3 | 0.1 | 1.8 |
| 30 | 39 | 80 | -13526 | 21928 | 320.6 | 0.0 | -0.6 |
| 31 | 40 | 305 | 37841 | 21918 | 320.3 | 0.1 | 1.7 |
| 32 | 40 | 319 | 29447 | 21918 | 320.3 | 0.1 | 1.3 |
| 34 | 40 | 253 | 38976 | 21901 | 319.8 | 0.1 | 1.8 |
| 35 | 41 | 299 | -317 | 21942 | 321.0 | 0.0 | 0.0 |
| 36 | 41 | 308 | 54570 | 21944 | 321.0 | 0.2 | 2.5 |
| 37 | 41 | 49 | 11896 | 21906 | 319.9 | 0.0 | 0.5 |
| 38 | 43 | 270 | 14312 | 21630 | 311.9 | 0.0 | 0.7 |
| 39 | 43 | 325 | 38832 | 21638 | 312.1 | 0.1 | 1.8 |
| 40 | 46 | 70 | -30341 | 22648 | 342.0 | -0.1 | -1.3 |
| 41 | 50 | 273 | 15436 | 21544 | 309.4 | 0.0 | 0.7 |
| 42 | 53 | 325 | 42043 | 21878 | 319.1 | 0.1 | 1.9 |
| 43 | 53 | 54 | 2248 | 21845 | 318.1 | 0.0 | 0.1 |
| 44 | 54 | 59 | -11568 | 21487 | 307.8 | 0.0 | -0.5 |